%% file: main.tex
\documentclass[runningheads]{llncs}
\usepackage{graphicx}
\usepackage{wrapfig}
\usepackage{blindtext}
\usepackage{enumerate}
\usepackage{amsmath}
\usepackage{amsfonts}
\usepackage{color}

\usepackage{listings}
\usepackage{xcolor}

\usepackage{array}
\newcolumntype{P}[1]{>{\centering\arraybackslash}p{#1}}

\definecolor{codegreen}{rgb}{0,0.6,0}
\definecolor{codegray}{rgb}{0.5,0.5,0.5}
\definecolor{codepurple}{rgb}{0.58,0,0.82}
\definecolor{backcolor}{rgb}{0.95,0.95,0.92}

\lstdefinestyle{mystyle}{
	backgroundcolor=\color{backcolor},
	commentstyle=\color{codegreen},
	keywordstyle=\color{magenta},
	numberstyle=\tiny\color{codegray},
	stringstyle=\color{codepurple},
	basicstyle=\ttfamily\scriptsize,
	breakatwhitespace=false,
	breaklines=true,
	captionpos=b,
	keepspaces=true,
	numbers=left,
	numbersep=5pt,
	showspaces=false,
	showstringspaces=false,
	showtabs=false,
	tabsize=2
}

\input{macros}

\begin{document}
\title{RTAEval : A framework for evaluating runtime assurance logic}
%
%
\author{Kristina Miller\inst{1}
\and
Christopher K. Zeitler \inst{2}\and
William  Shen \inst{1}\and
Mahesh Viswanathan \inst{1} \and
Sayan Mitra \inst{1}}
\authorrunning{K. Miller et al.}
%
\institute{University of Illinois Urbana Champaign, Champaign IL 61820, USA \and
Rational CyPhy, Inc., Urbana IL 62802}
\maketitle              
\begin{abstract}
\input{sections/abstract}
\end{abstract}

\input{sections/intro}

\input{sections/overview}
\input{sections/conclusion}

\bibliographystyle{splncs04}
\bibliography{refs}

\end{document}

%% file: macros.tex
\newcommand{\toolname}{\textsf{RTAEval}}

\newcommand{\simplesim}{\textsf{simpleSim}}

\newcommand{\statespace}{\mathcal{X}}

\newcommand{\unsafe}{\mathcal{O}}

\newcommand{\nnreals}{\mathbb{R}_{\geq 0}}

\newcommand{\safety}{\texttt{S}}
\newcommand{\untrusted}{\texttt{U}}
\newcommand{\lead}{L}

\newcommand{\step}{\texttt{step}}
\newcommand{\updateDef}{\texttt{updateDef}}
\newcommand{\rtaswitch}{\texttt{RTASwitch}}
\newcommand{\rtalogic}{\texttt{RTALogic}}
\newcommand{\collectTrace}{\texttt{collect\_trace}}
\newcommand{\collectCompTimes}{\texttt{collect\_computation\_times}}

\newcommand{\cstv}{\bar{v}}

\newcommand{\reachrta}{\textsf{ReachRTA}}

\newcommand{\simrta}{\textsf{SimRTA}}

\newcommand{\acc}{\textsf{ACC}}
\newcommand{\dubins}{\textsf{Dubins}}
\newcommand{\acas}{\textsf{GCAS}}

%% file: sections/abstract.tex
Runtime assurance (RTA) addresses the problem of keeping an autonomous system safe while using an untrusted (or experimental) controller.
This can be done via logic that explicitly switches between the untrusted controller and a safety controller, or logic that filters the input provided by the untrusted controller.
While several tools implement specific instances of 
RTAs,
there is currently no framework for evaluating different approaches. Given the importance of the RTA problem in building safe autonomous systems, an evalutation tool is needed. 
In this paper, we present the $\toolname$ framework 
as a low code framework that can be used
to quickly evaluate different RTA logics for different types of agents in a variety of scenarios.
$\toolname$ is designed to quickly create scenarios, run different RTA logics, and collect data that can be used to evaluate and visualize  performance. In this paper, we describe different components of $\toolname$ and show how it can be used to create and evaluate scenarios involving multiple aircraft models.


\keywords{Runtime assurance  \and  Autonomous systems.}

%% file: sections/intro.tex
\section{Introduction}
As autonomous systems are deployed in the real world, their safe operation is becoming critical in a number of domains such as  aerospace, manufacturing, and transportation. The need for safety is often at odds with the need to experiment with, and therefore deploy, new untrusted technologies in the public sphere. 
For example, experimental  controllers created using  reinforcement learning can provide  better performance in simulations and controlled environments, but assuring safety in real world circumstances is currently beyond our capabilities for such controllers. 
{\em Runtime assurance (RTA)\/}~\cite{Simplex-Sha,sha2001using,bak2011sandboxing,RTA-VV-SC-FCS-2008} addresses this tension. The  idea is to introduce a  {\em decision module\/} that somehow chooses between a well-tested {\em Safety controller\/} and the experimental, {\em Untrusted controller\/}, assuring safety of the overall system while also allowing experimentation with the new untrusted technology where and when possible. 
Specific RTA  technologies are being researched and  tested  for aircraft engine control~\cite{RTA-AFC2010}, air-traffic management~\cite{cofer-rta-22}, and satellite  rendevous and proximity operations~\cite{dunlap2023run}.

The Simplex architecture~\cite{Simplex-Sha,sha2001using} first proposed this idea in a form that is recognizable as RTA.
Since then, the central problem of designing a {\em decision module\/}  that chooses  between the  different controllers has been addressed in a  number of works such as SimplexGen~\cite{bak2011sandboxing}, Black-Box Simplex~\cite{mehmood2022black}, and SOTER~\cite{desai2019soter}.
The two main approaches for building the decision module are based on
(a) an RTASwitch which chooses  one of the controllers using the current state
 or (b) an RTAFilter which blends the outputs from the two controllers to create the final output. 
In creating an RTASwitch, the decision can be based on forward-simulation of the current state~\cite{wadley2013development}, model-based~\cite{bak2011sandboxing} and model-free forward reachability~\cite{mehmood2022black}, or model-based backward reachability~\cite{bak2011sandboxing}.
%
The most common filtering method is Active Set Invariance Filtering (ASIF)~\cite{ames2019control}, wherein a control barrier function is used to blend the control inputs from the safety and untrusted controllers such that the system remains safe with respect to the control barrier functions~\cite{hibbard2022guaranteeing,mote2021natural,dunlap2022run}.




%

While these design methods for the decision module have evolved quickly, a software framework for evaluating the different techniques has been missing.
In this paper, we propose such a flexible, low-code framework called $\toolname$ (Figure~\ref{fig:rta-framework}).
The  framework consists of a module for defining scenarios, possibly involving multiple agents; a module for executing the defined  scenario with suitable RTASwitches and RTAFilters; and a module for collecting and visualizing  execution data. RTAEval  allows different agent dynamics, decision modules, and metrics to be plugged-in with a few lines of code. 
%
In creating RTAEval, we have defined standardized interfaces 
between the agent simulator, the decision module (RTA), and data collection.

In Section~\ref{sec:framework}, we give an overview of  $\toolname$.
In Section~\ref{sec:scenarios}, we discuss how scenarios are defined, 
and, in  Section~\ref{sec:rta}, we discuss how the user should provide decision modules (also called the RTALogic).
%
In Section~\ref{sec:eval}, we discuss  data collection, evaluation, and visualization.
Finally, in Section~\ref{sec:examples}, we  show a variety of examples implemented in $\toolname$.
A tool suite for this framework can be found at \url{https://github.com/RationalCyPhy/RTAEval}.

%% file: sections/overview.tex
\section{Overview of the $\toolname$ Framework}
\label{sec:framework}

\begin{figure}[ht!]
	\centering	
	\includegraphics[width=\linewidth]{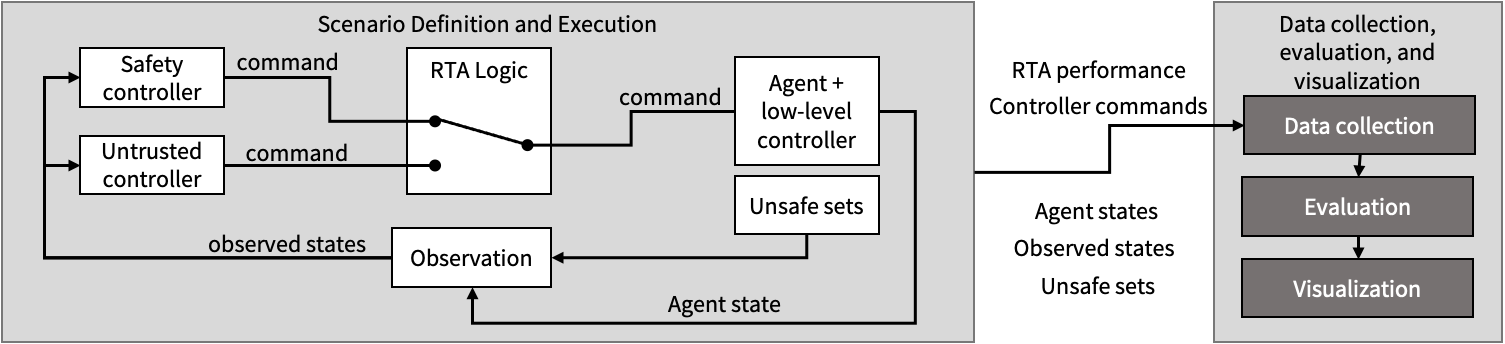}
	\caption{\scriptsize $\toolname$ framework. Some user-defined scenario is executed, and the execution data is collected, evaluated, and visualized using our provided suite of tools. The scenario is defined by the safety controller, untrusted controller, plant and low level controllers, unsafe sets, and sensor.}
	\label{fig:rta-framework}
\end{figure}

The three main components of $\toolname$ are (a) the scenario definition, (b) the scenario execution, and (c) the data collection, evaluation, and visualization module (See Figure~\ref{fig:rta-framework}).
A scenario is defined by the agent and its low-level controller, the unsafe sets, the untrusted and safety controllers, the time horizon for analysis, and the initial conditions.
Given this scenario definition, the scenario is executed iteratively over the specified time horizon.

During each iteration of the closed-loop execution of the RTA-enabled autonomous system, 
the current state of the agent and the sets of unsafe states are collected. 
This observed state information is given to both the untrusted and safety controller, which each compute control commands.
Both of these commands are evaluated by the user-provided decision module (i.e., RTA logic), which computes and returns the actual command to be used by the agent.
The agent then updates its state, and the computation moves to the next iteration.
While the execution proceeds, data $\textendash$  such as the RTA computational performance, controller commands, agent states, and observed state information of the unsafe sets $\textendash$
is collected via the data collection module. At the end of an execution, this data is evaluated to summarize the overall performance of the RTA. This summary includes computation time of the RTA logic, untrusted versus safety controller usage, and the agent's distance from the unsafe set.
We also provide a visualization of this data.

A low-code tool suite of the $\toolname$ framework is written in Python, and it is flexible in that it allows for a wide variety of simulators and can be generalized to scenarios where multiple agents are running a variety of different RTA modules. Simple Python implementations of vehicle models (some of which we provide in $\simplesim$) can be incorporated directly.
However, users can incorporate new agent models within $\simplesim$ as long as the agent has a function $\step$ that defines the dynamics and low-level controller of the agent and returns the state of the agent at the next time step. An example of this is provided in Example~\ref{ex:acc-def} and Figure~\ref{fig:acc-step}. The safety and untrusted controllers should also be encoded in $\step$, which simply takes in the command (or mode) to be used over the next time step.  
Higher fidelity simulators such as CARLA~\cite{dosovitskiy2017carla} and AirSim~\cite{shah2018airsim} can also be used in place of $\simplesim$ for the execution block. 
The observed state information would need to be provided to our data collection, evaluation, and visualization tool in the format seen in Figure~\ref{fig:exec_dict}.
%
%

\vspace{-0.75cm}
\subsection{Scenario definition and execution}
\label{sec:scenarios}
A {\em scenario} is defined by the {\em agent}, {\em unsafe sets}, {\em safety} and {\em untrusted controllers}, {\em initial conditions}, and {\em time horizon} $T > 0$.
The {\em simulation state at time $t \in [0,T]$} consists of the {\em agent state}, the {\em unsafe set definition}, and the {\em control command} at time $t$.
The agent has an identifier, a state, and some function $\step$ that takes in some control command at time $t$ and outputs the system state at time $t+1$.
The {\em unsafe sets} are the set of states that the system must avoid over the execution of the scenario.
We say that the agent is {\em safe} if it is outside the unsafe set.
The safety and untrusted controllers compute control commands for the system, which are then filtered through the RTA logic, as discussed further in Section~\ref{sec:rta}.
The initial conditions define the simulation state at time $0$.
Then, given a scenario with some time horizon $T$ and an RTA logic, an {\em execution} of the scenario is a sequence of time-stamped simulation states over $[0,T]$.
Note that, while we define an execution as a discrete time sequence of simulation states, the actual or real-world execution of the scenario may be in continuous time;
thus, we simply sample the simulation states at a predefined interval.
We call the part of the execution that contains only the sequence of agent states the {\em agent state trace}.
Similarly, we call the part of the execution that only contains the sequence control commands the {\em mode trace} and the part that only contains the sequence of unsafe set states the {\em unsafe set state trace}.

\begin{wrapfigure}[14]{r}{0.55\textwidth}
    \centering
    \vspace*{-0.2in}
    \includegraphics[width=\linewidth]{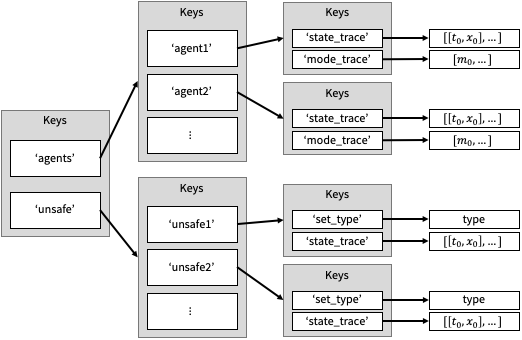}
    \caption{\scriptsize Execution structure required by the $\toolname$ evaluation and visualization}
    \label{fig:exec_dict}
\end{wrapfigure}

In order for our evaluation and visualization to work, the execution must be given to the data collection as a dictionary, the structure of which is shown in Figure~\ref{fig:exec_dict}.
Here, there are three levels of dictionaries.
The highest level dictionary has the keys $\texttt{`agents'}$ and $\texttt{`unsafe'}$, which point to dictionaries containing the state and mode traces of the agents and state traces of the unsafe sets respectively.
The second level of dictionaries has keys that correspond to different agents and unsafe sets.
We call these keys the agent and unsafe set IDs.
Each agent ID points to a dictionary containing the state and mode traces of that agent.
The state trace is a list of time-stamped agent states, and the mode trace is a sequential list of control commands.
Each unsafe set ID points to a dictionary containing the set type and state trace of that unsafe set.
The set type is a string that tells $\toolname$ what type of set that particular unsafe set is.
Currently, $\toolname$ supports the following set types: {\em point}, {\em ball}, {\em hyperrectangle}, and {\em polytope}.
Each set has a {\em definition} that, together with the type, defines the set of states contained within the unsafe set.
Then, the state trace for an unsafe set is a sequence of time stamped definitions of the set.

\begin{wrapfigure}[10]{l}{0.5\textwidth}
    \centering
    \vspace*{-0.2in}
    \includegraphics[width=\linewidth]{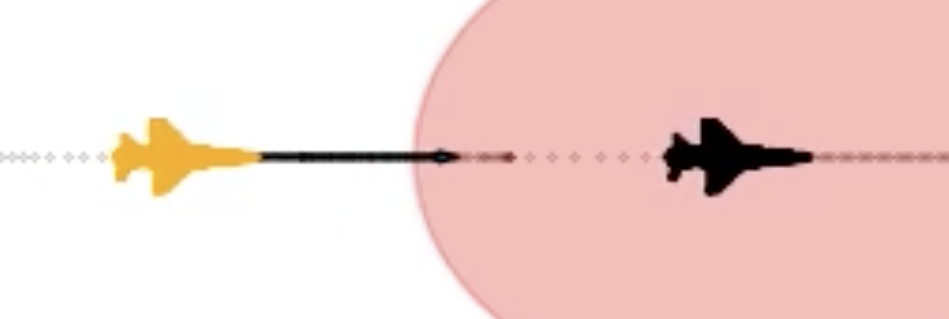}
    \caption{\scriptsize Example visualization of the scenario defined in Example~\ref{ex:acc-def}. The leader is shown in black, the follower is shown in orange, and the unsafe region is shown in red.}
    \label{fig:acc-fig}
\end{wrapfigure}

\begin{example}
\label{ex:acc-def}
Consider the following adaptive cruise control (ACC) scenario shown in Figure~\ref{fig:acc-fig} as a running example:
An agent with
state $x = [p, v]^{\top}$ has dynamics  given by
$$
f(x, m) = \begin{bmatrix}
    0 & 1 \\ 0 & 0
\end{bmatrix} x + \begin{bmatrix}
    0 \\ 1
\end{bmatrix} a(x, m),
$$
where $a(x, m) = g_{\safety}(x)$ if $m = \safety$ 
and $a(x, m) = g_{\untrusted}(x)$ if $m = \untrusted$.
The agent tries to follow at distance $d > 0$ behind a leader moving at constant speed $\cstv$.
The position of the leader at time $t$ is given by $p_{\lead}(t)$.
Then, the untrusted controller $g_{\untrusted}$ and safety controller $g_{\safety}$ are given by 
$$
    g_{\safety}(x) = k_1 ((p_{L}(t) - d) - p) + k_2 (\cstv - v) \textrm{ and } g_{\untrusted}(x) = \begin{cases}
        a_{\max} & (p_{L}(t) - p) > d \\
        -a_{\max} & \textrm{else}
    \end{cases},
$$
where $k_1 > 0$ and $k_2 > 0$.
The function $\step$ is a composition of the untrusted controller, the safety controller, and the dynamics function of the system.

A collision between the agent and leader occurs if $\|p_{\lead}(t) - p(t)\| \leq c$, $c < d$.
There is then an unsafe set centered on the leader agent,
and it is defined by 
$\unsafe = \{ [p, v, t]^{\top} \in \statespace \times \nnreals \ | \ \|p_{\lead}(t) - p \| \leq c \}$.
The function $\updateDef$ then takes in the current state of the simulator and creates the unsafe set centered on the leader.
The initial conditions for this scenario are then the initial agent state $x_0$, the initial leader state $x_{\lead0}$,
and the time horizon $T > 0$.
\end{example}

This scenario is shown in our low code framework in Figures~\ref{fig:acc-step} and~\ref{fig:simpleSim}.
The dynamics of the agent are defined in $\step$ in lines 11-26 of Figure~\ref{fig:acc-step}.
The proportional controller is defined in lines 1-4 and the bang-bang controller is defined in lines 6-9.
This is all contained within a class $\texttt{AccAgent}$.
In Figure~\ref{fig:simpleSim}, we set up the scenario.
In lines 2-5, we define the goal point for the agent.
In lines 7-15, we create the agent, the leader, and the unsafe set.
Finally, in line 18, we initialize the scenario to be executed; in lines 21-26, we add the agents and unsafe sets to the scenario; and in lines 29-30, we set up the scenario parameters.


\begin{figure}[ht!]
\begin{lstlisting}[language=Python]
def P(self, time_step): #Proportional controller
    xrel = self.goal_state[0] - self.state_hist[-1][0]
    vrel = self.goal_state[1] - self.state_hist[-1][1]
    return self.Kp[0]*xrel + self.Kp[1]*vrel

def BangBang1D(self, time_step): # Bang-bang controller
    x_err_curr = self.goal_state[0] - self.state_hist[-1][0]
    v_err_curr = self.goal_state[1] - self.state_hist[-1][1]
    return np.sign(x_err_curr)*self.a_max

def step(self, mode, initialCondition, time_step, simulatorState):
        self.goal_state = self.desired_traj(simulatorState)
        if mode == 'SAFETY':
            self.control = self.P
        elif mode == 'UNTRUSTED':
            self.control = self.BangBang1D
        else:
            self.control = self.no_control()
        a_curr = self.control(time_step)
        if abs(a_curr) > self.a_max:
            a_curr = np.sign(a_curr)*self.a_max
        x_next = initialCondition[0] + initialCondition[1]*time_step
        v_next = initialCondition[1] + a_curr*time_step
        if abs(v_next) >= self.v_max:
            v_next = np.sign(v_next)*self.v_max
        return [x_next, v_next]
\end{lstlisting}
\caption{\scriptsize Controller and step functions for the agent in Example~\ref{ex:acc-def}. The first function defined is the proportional controller (Safety) and the second function is the Bang-bang controller (Untrusted). The step function (lines 11-26) takes in the current mode and state of the agent, as well as the time stpe and current suimulator state. In lines 13-18 it decides which controller to use, and in lines 19-26, it updates the staet of the agent.}
\label{fig:acc-step}
\end{figure}

\begin{figure}[ht!]
\begin{lstlisting}[language=Python]
# Define desired goal point for the follower agent (ego):
def agent1_desiredTraj(simulationTrace):
    lead_state = simulationTrace['agents']['leader']['state_trace'][-1][1:]
    return [lead_state[0] - 10, lead_state[1]]
    
# Create the ego agent:
agent1 = AccAgent("follower",file_name=controllerFile)
agent1.follower = True
agent1.desired_traj = agent1_desiredTraj

# Create the leader agent:
leader = AccAgent("leader",file_name=controllerFile)

# Create the unsafe set centered on leader agent:
unsafe1 = relativeUnsafeBall("unsafe1", [5], 7, "leader")

# Initialize the ACC scenario:
accSim = simpleSim()

# Initialize agents and unsafe sets in the scenario:
agents = [agent1, leader]
RTAs = [None, None]
modes = [ccMode.UNTRUSTED, ccMode.NORMAL]
inits = [[0,1], [5,1]]
accSim.addAgents(agents=agents, modes=modes, RTAs=RTAs, initStates=inits)
accSim.addUnsafeSets(unsafe_sets = [unsafe1])

# Set simulation parameters
accSim.setSimType(vis=False, plotType="2D", simType="1D")
accSim.setTimeParams(dt=0.1, T=5)
\end{lstlisting}
\caption{\scriptsize
Python code snippet defining the scenario in our low-code $\toolname$ framework.
The untrusted and safety controllers are contained within the dynamics of the $\texttt{AccAgent}$, which is defined in a separate file.
The scenario is executed in a simply python simulator, which is initiated on line 18.
Initially, the agents are not assigned RTAs, but this will be done in Section~\ref{sec:rta}.
The agents and unsafe sets are added to the scenario, and the simulation parameter are set in lines 29 and 30.
}
\vspace{-5.1mm}
\label{fig:simpleSim}
\end{figure}

\subsection{RTA logics}
\label{sec:rta}
We provide an RTA base class that can be used in $\toolname$. The user must  provide the RTA logic to be evaluated.
This logic takes in an observed state  and outputs the control command to be used by the plant.
This observed state information has to be  provided in the format shown in Figure~\ref{fig:exec_dict} for data collection, evaluation, and visualization to work. The RTA base class is shown in Figure~\ref{fig:base-rta}.
We provide the functions $\rtaswitch$ and $\texttt{setupEval}$.
Users must provide the switching logic as $\rtalogic$.
When creating RTA, the user can decide to use our data collection by running $\texttt{setupEval}$ in $\texttt{\_\_init\_\_}$.
This will create a data collection object called $\texttt{eval}$, which saves the data used for our evaluation (see Section~\ref{sec:eval}).
The switch is performed in $\rtaswitch$, which also stores the current perceived state of the simulator from the point of view of the agent, as well as the time to compute the switch.
The user provided  switching logic $\rtalogic$  takes in the current state of the simulator and returns the mode that the agent should operate in.
To create different logics, the user must create an RTA class derived from the  RTA base class, which  implements the function $\rtalogic$.
An example of this is given in Example~\ref{ex:acc-rta}.

\begin{figure}[ht!]
\begin{lstlisting}[language=Python]
class baseRTA(abc.ABC):
    def __init__(self):
        self.do_eval = False # Don't automatically set up RTAEval
        pass

    @abc.abstractmethod
    def RTALogic(self, simulationTrace: dict) -> Enum:
        # User provided logic for switching RTA
        pass

    def RTASwitch(self, simulationTrace: dict) -> Enum:
        start_time = time.time()
        rtaMode = self.RTALogic(simulationTrace)
        running_time = time.time() - start_time

        if self.do_eval:
            self.eval.collect_computation_time(running_time)
            self.eval.collect_trace(simulationTrace)
        return rtaMode

    def setupEval(self): 
        # Add this to init when inheriting baseRTA to inclde evaluation
        self.do_eval = True
        self.eval = RTAEval()
\end{lstlisting}
\caption{\scriptsize Base RTA class used in the low-code $\toolname$ framework. Users need only provide the decision logic, which we call $\rtalogic$.}
\label{fig:base-rta}
\end{figure}

\begin{example}
    \label{ex:acc-rta}
    An example of a simple RTA switching logic can be seen in Figure~\ref{fig:rta-example}.
    This is a simulation-based switching logic that was designed for the adaptive cruise control introduced in Example~\ref{ex:acc-def}.
    Here, the future states of the simulator are predicted over some time horizon $T$ and saved as $\texttt{predictedTraj}$ in line 2.
    We then check over this predicted trajectory to see if the agent ever enters the unsafe set in lines 3-11.
    If it does, then the safety controller is used, and if it does not, then the untrusted controller is used.
    Once $\rtalogic$ is created, we add it to a new class called $\texttt{accSimRTA}$ and use it to create an RTA object called $\texttt{egoRTA}$ for $\texttt{egoAgent1}$.
    We can then change line~22 in Figure~\ref{fig:simpleSim} to $\texttt{RTAs = [egoRTA, None]}$.
    This will associate $\texttt{egoRTA}$ with $\texttt{egoAgent1}$ and run the RTA switching logic every time the state of $\texttt{egoAgent1}$ is updated.
\end{example}

\begin{figure}[ht!]
\begin{lstlisting}[language=Python]
def RTALogic(simulationTrace):
    predictedTraj = simulate_forward(simulationTrace)
    egoTrace = predictedTraj['agents'][egoAgent.id]['state_trace'] 
    for unsafeSet in self.unsafeSets:
        unsafeSetTrace = predictedTraj['unsafe'][unsafeSet.id]['state_trace']
        for i in range(len(unsafeSetTrace)):
            egoPos = egoTrace[i][1]
            unsafeSetDef = unsafeSetTrace[i][1]
            pos_max = unsafeSetDef[0][0] - unsafeSetDef[1]
            if egoPos > pos_max:
                return egoModes.SAFETY
    return egoModes.UNTRUSTED
\end{lstlisting}
\caption{\scriptsize Example RTA switching logic for Example~\ref{ex:acc-rta}. Here, the trajectory of the follower agent is simulated forward, and if it ever comes within collision distance of the leader, then the safety controller is used.}
\label{fig:rta-example}
\end{figure}

\subsection{Data collection, evaluation, and visualization}
\label{sec:eval}
We now discuss the data collection, evaluation, and visualization tool which is provided as a part of $\toolname$.
This tool is a class that has some collection functions and post-processing functions.
To use the data collection and evaluation functionalities provided, the user must add the line $\texttt{self.setupEval()}$ when creating the RTA object.
Data collection occurs via the functions $\collectTrace$ and $\collectCompTimes$.
Here, $\collectTrace$ collects the simulation traces, and $\collectCompTimes$ collects the time it takes for the RTA module to compute a switch.
An example of how the data collection can be incorporated in the RTA module is shown in Figure~\ref{fig:rta-example}.
The traces are collected and stored as a dictionary of the form shown in Figure~\ref{fig:exec_dict}. 
%
Once the data has been collected over a scenario, we can use them to evaluate the performance of the RTA over a scenario.
Examples of the data evaluation, as well as screenshots from our simulator are shown in Section~\ref{sec:examples}.
A summary of the RTA's performance in the scenario can be quickly given by running $\texttt{eval.summary()}$.
The main metrics that we study are the following:
{\bf Computation time} gives the running time of $\rtaswitch$ each time it is invoked.
We provide the average, minimum, and maximum times to compute the switch.
{\bf Distance from unsafe set} is the distance between the ego agent and the unsafe sets. We also allow the user to find the distance from other agents in the scenario.
{\bf Time to collision (TTC)} is the time until collision between the ego agent and the other agents if none of them change their current trajectories.
Finally, we also provide information on the {\bf percent controller usage}, which is the proportion of time each controller is used over the course of the scenario.
We also provide information on the number of times a switch occurs in a scenario.
Example results are shown in Section~\ref{sec:examples}.

\section{$\toolname$ Examples}
\label{sec:examples}
\begin{figure}[ht!]
    \centering
    \includegraphics[width=0.35\linewidth]{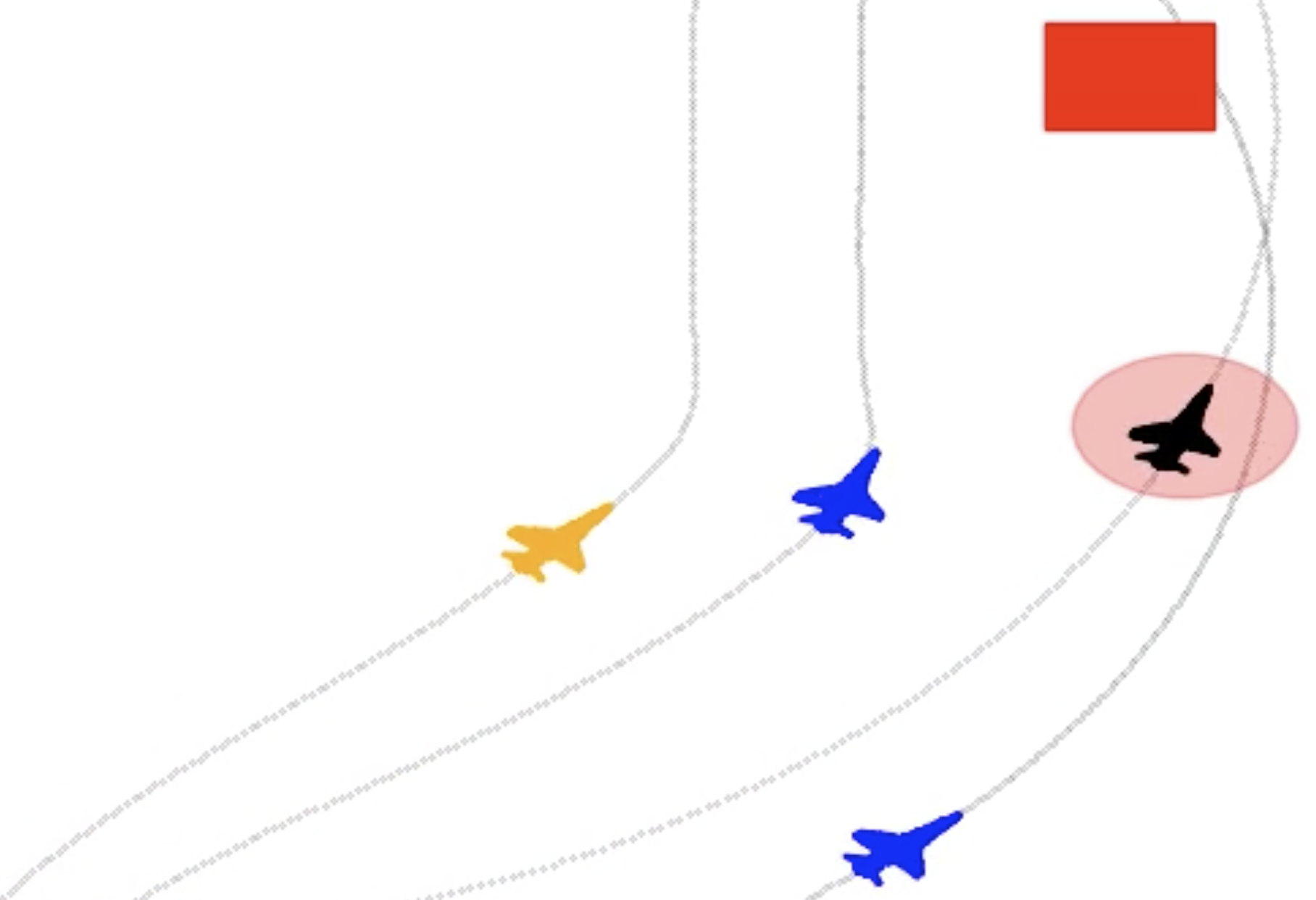}
    \includegraphics[width=0.6\linewidth]{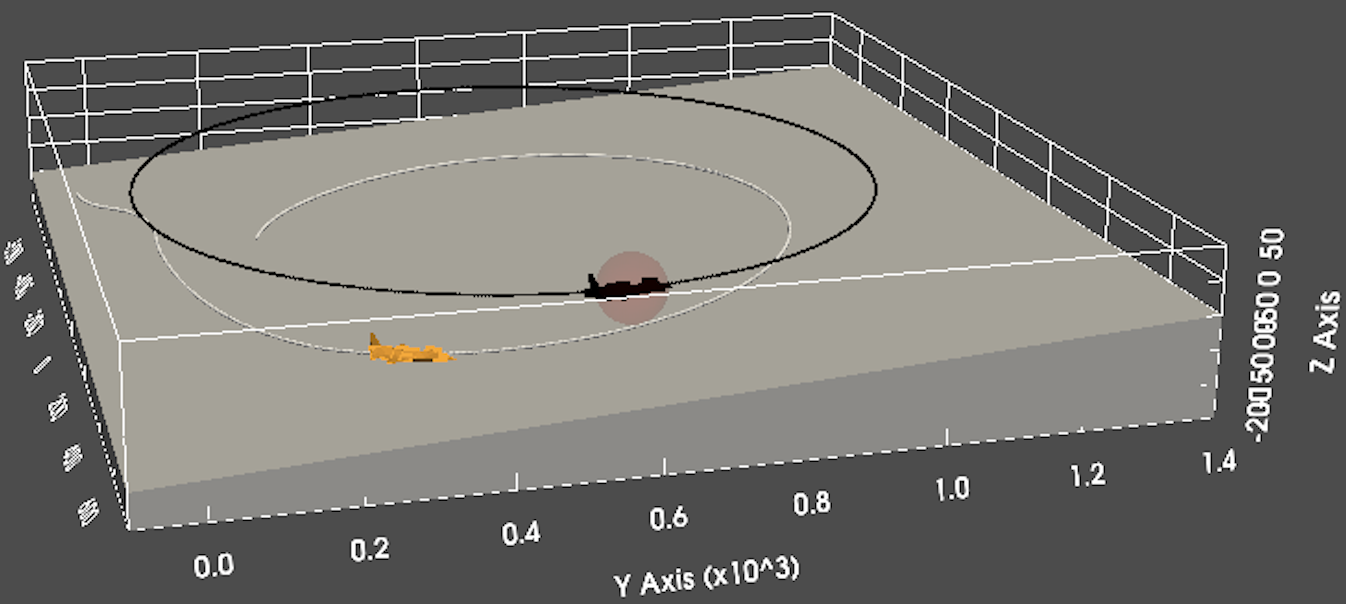}
    \caption{\scriptsize Example scenarios in Table~\ref{tab:scenarios}. Left: 2-dimensional dubins aircraft with building collision avoidance. The leader is shown in black, and the followers are shown in orange and blue. The desired trajectories are shown in gray. Right: Ground collision avoidance. The leader is shown in black and the follower is shown in orange. The desired trajectory is shown in white.}
    \label{fig:dubins}
\end{figure}

In this section, we present some examples using our provided suite of tools for $\toolname$.
We evaluate two different decision module logics: $\simrta$ and $\reachrta$.
$\simrta$ is the simulation based switching logic introduced in Example~\ref{ex:acc-rta}.
$\reachrta$ is similar to $\simrta$ but uses reachable sets that contain all possible trajectories of the agent as the basis of the switching logic.
%
We evaluate these RTAs in $1$-, $2$-, and $3$-dimensional scenarios with varying numbers of agents.
These scenarios are described in more detail in Table~\ref{tab:scenarios}.
Here, the workspace denotes the dimensions of the physical space that the systems live in. 
Note that, while all the examples presented have some physical representation, this is not a necessary requirement of the tool.
We also provide pointers to where the dynamics of the agents can be found, as well as the untrusted and safety controllers used.
Visualizations of the scenarios can be seen in Figures~\ref{fig:acc-fig} and~\ref{fig:dubins}.

\begin{table}[ht!]
    \centering
    \begin{tabular}{p{0.18\linewidth}||P{0.25\linewidth}|P{0.25\linewidth}|P{0.25\linewidth}}
        & $\acc$ & $\dubins$ & $\acas$ \\ \hline \hline
        Workspace & $1$ & $2$ & $3$ \\ \hline
        Dynamics & Example~\ref{ex:acc-def} & Dubin's car~\cite{dunlap2021comparing} & Dubin's plane~\cite{dunlap2021comparing} \\
        Untrusted & Bang-bang controller (Example~\ref{ex:acc-def}) & PID with accleration~\cite{fan2020fast} & PID with acceleration~\cite{fan2020fast}\\ \hline
        Safety & PID (Example~\ref{ex:acc-def}) & PID with deceleration~\cite{fan2020fast} & PID with deceleration and pitching up~\cite{fan2020fast} \\ \hline
        Unsafe & Leader (ball) & Leader (ball) and building (rectangle) & Leader (ball) and ground (polytope) \\ \hline
        Visualization & Figure~\ref{fig:acc-fig} & Figure~\ref{fig:dubins} & Figure~\ref{fig:dubins} \\ \hline
        Scenario length & 10 s & 20 s & 40 s \\ \hline
    \end{tabular}
    \caption{\scriptsize Brief description of evaluated scenarios.}
    \label{tab:scenarios}
\end{table}

Each of these scenarios is executed using $\simplesim$, and the three RTA logics are created for them.
Data is collected over the scenario lengths in Table~\ref{tab:scenarios}.
Note that the scenario length is the simulation time for the scenario and not the real time needed to run the scenario.
We run these scenarios with varying numbers of agents and present the running time of the scenario execution and evaluations in Table~\ref{tab:runningtimes}.
The simulation time step is set to $0.05$ for all scenarios.
Here, exec time is the time it takes to run the scenario, RTA comp time is the the average time it takes for the user provided RTA logic to make a decision,
\% RTA comp is the percentage of the exec time that is taken by the RTA decision module, and eval time is the time it takes to get a full summary of how the RTA performs for each agent.
The evaluation summary includes the average decision module computation time, controller usage, distance from the unsafe sets and other agents, and time to collision with the unsafe sets and other agents.
We note that a majority of the run time for the scenario execution is due to the RTA logic computation time and not our tool.
Additionally, while the run time of the evaluation is 
affected by the number of agents in the scenario, it is
mostly affected by the set type of the unsafe set,
where the polytope in the $\acas$ scenario causes the biggest slow down in evaluation time.

\begin{table}[ht!]
    \centering
    \begin{tabular}{c|c||c|c|c|c||c|c|c|c}
        \multicolumn{2}{c||}{}& \multicolumn{4}{c||}{SimRTA} & \multicolumn{4}{c}{ReachRTA} \\ \hline
        Scenario & \begin{tabular}[c]{@{}c@{}}Num\\ agents\end{tabular} & \begin{tabular}[c]{@{}c@{}}Exec\\ time\end{tabular} & \begin{tabular}[c]{@{}c@{}}RTA comp\\ time (ms)\end{tabular} & \begin{tabular}[c]{@{}c@{}}\% RTA\\ Comp\end{tabular} & \begin{tabular}[c]{@{}c@{}}Eval\\ time\end{tabular} & \begin{tabular}[c]{@{}c@{}}Exec\\ time\end{tabular} & \begin{tabular}[c]{@{}c@{}}RTA comp\\ time (ms)\end{tabular} & \begin{tabular}[c]{@{}c@{}}\% RTA\\ Comp\end{tabular} & \multicolumn{1}{c}{\begin{tabular}[c]{@{}c@{}}Eval\\ time\end{tabular}} \\ \hline \hline
        ACC & 1 &
        18.49e-3 & 0.07 & 76.63 & 7.27e-3 &
        0.35 & 1.71 & 97.89 & 8.22e-3 \\
            & 2 &
        50.08e-3 & 0.10 & 84.12 & 18.16e-3 &
        1.12 & 2.76 & 98.66 & 17.96e-3 \\
            & 5 &
        0.18 & 0.16 & 90.47 & 87.96e-3 & 
        6.01 & 5.96 & 99.10 & 0.10 \\ \hline
        Dubins & 1 & 
        2.32 & 4.99 & 86.06 & 32.88e-3 & 
        15.18 & 37.10 & 97.76 & 34.15e-3 \\
               & 3 & 
        15.30 & 11.84 & 92.89 & 0.18 & 
        71.60 & 58.83 & 98.60 & 0.11 \\
               & 10 &
        203.87 & 49.77 & 97.65 & 0.70 &
        461.71 & 114.08 & 98.83 & 0.76 \\ \hline
        GCAS & 1 &
        5.85 & 6.08 & 83.12 & 30.62 &
        39.28 & 47.65 & 97.02 & 31.423 \\
             & 1 &
        45.27 & 17.84 & 94.60 & 83.61 &
        174.00 & 71.11 & 98.07 & 98.10 \\ \hline
\end{tabular}
    \caption{\scriptsize Running time for execution and evaluation of RTAs with the tool suite provided for RTAEval. The times are given in seconds unless otherwise indicated.}
    \label{tab:runningtimes}
\end{table}

\begin{figure}[ht]
    \centering
    \includegraphics[width = 0.32 \linewidth]{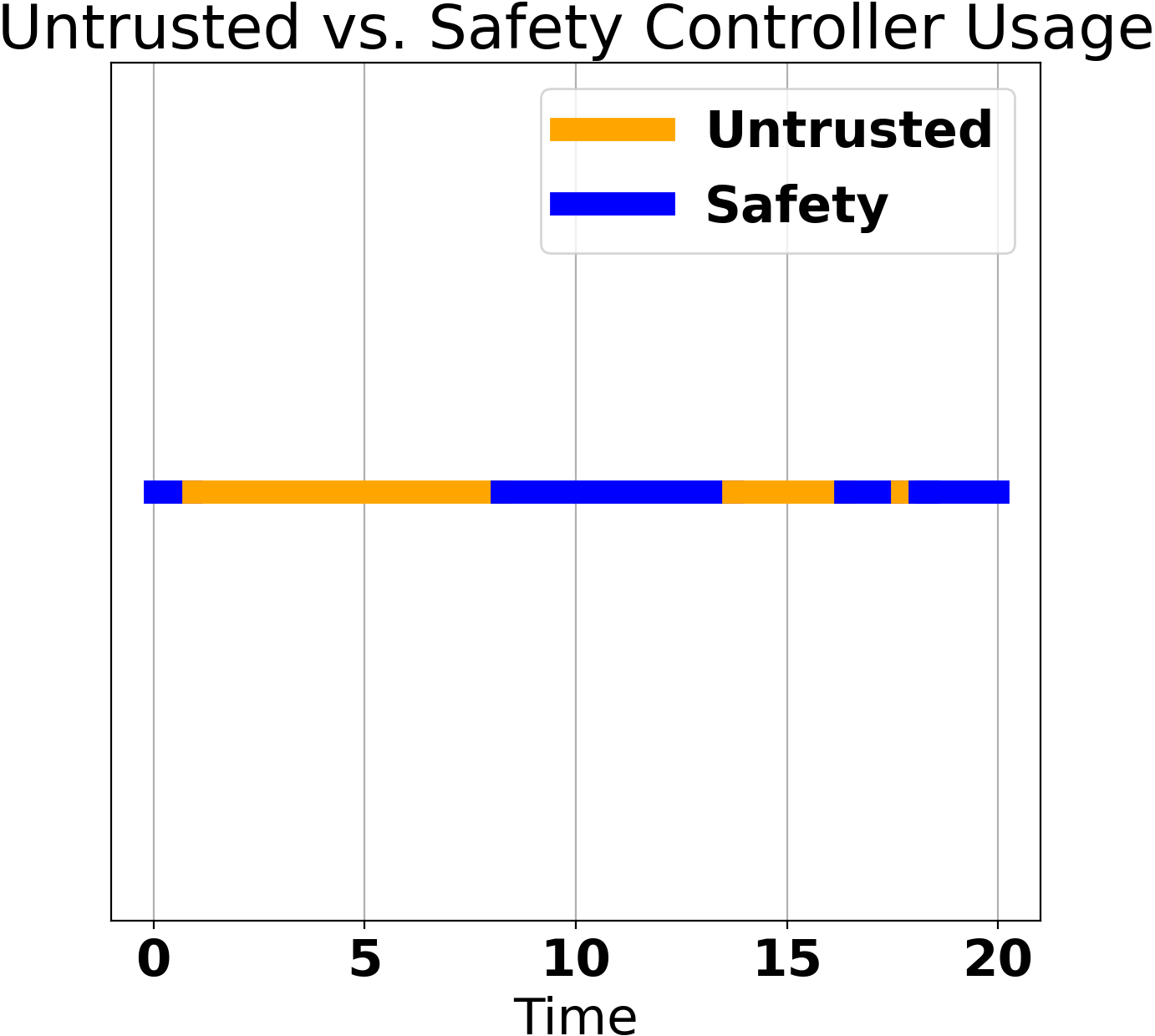}
    \includegraphics[width = 0.27 \linewidth]{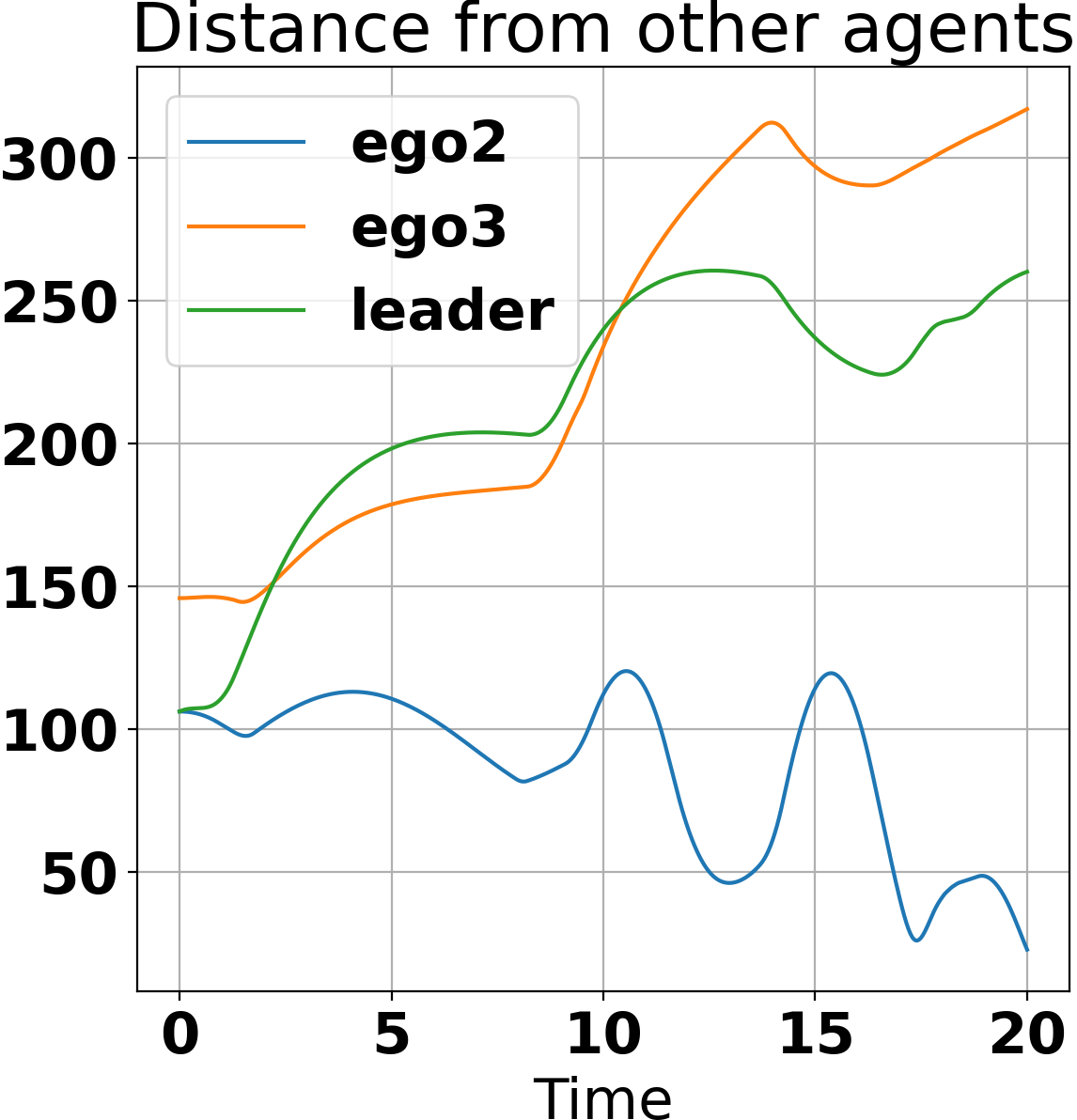}
    \includegraphics[width = 0.32 \linewidth]{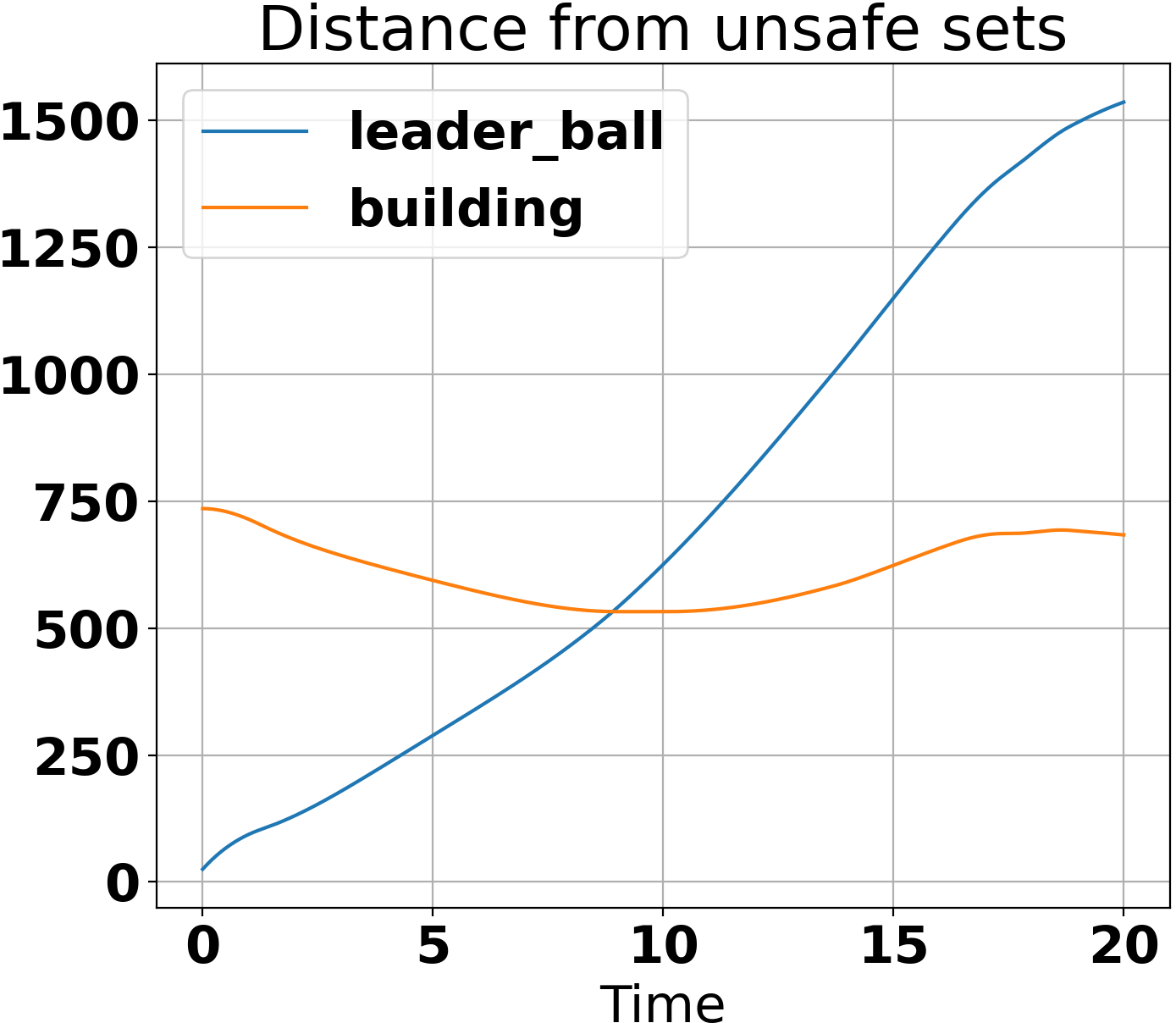}
    \caption{ \scriptsize Example visualization from three agent dubins scenario.
    From left to right: Controller usage plot, distance from other agents, and distance from unsafe sets. \textbf{ego2} and \textbf{ego3} denote the other aircraft.}
    \label{fig:ex-vis}
\end{figure}

The summary of an RTA performance is given out in a text file from which visualizations like the one in Figure~\ref{fig:ex-vis} can be easily created.
In addition to the computation time, distance from the unsafe sets, distance from the other agents, and controller usage, the minimum times to collision (TTC) for the unsafe sets and other agents are also reported.
The summary information is saved in such a way that users can pull up snapshots of the scenario at any point in time. 
This means that the user can examine the state of the scenario that caused an unwanted result.
Such functionality aids in the rapid prototyping of RTA technologies and logics.








%% file: sections/conclusion.tex
\vspace{-0.75cm}
\section{Conclusion}
We presented the $\toolname$  suite of Python-based tools for evaluating different runtime assurance (RTA) logics. Different RTA switching logics can be quickly coded in $\toolname$,
and we demonstrate its functionality in rapid prototyping of RTA logics on a variety of examples.
$\toolname$ can be used in  multi-agent scenarios and scenarios with  perception models.
Interesting next steps might include extending the functionality of $\toolname$ to filtering methods such as ASIF and scenarios that involve effects of proximity-based communication.  